\documentclass[openacc]{rstransa}

\input{userDefs/myPreamble}
\input{userDefs/myCommands}

\begin{document}

\title{Routes to turbulence in \TCF}

\author{Daniel Feldmann$^1$, Daniel~Borrero-Echeverry$^{2}$, Michael~J.~Burin$^{3}$, Kerstin~Avila$^{4,5}$ and Marc~Avila$^{1,6}$}

\address{%
$^{1}$University of Bremen, Center of Applied Space Technology and Microgravity (ZARM), 
28359 Bremen, Germany.\\
$^{2}$Department of Physics, Willamette University, 
Salem, OR 97301, USA.\\
$^{3}$Department of Physics, California State University - San Marcos, 
San Marcos, CA 92096, USA.\\
$^{4}$University of Bremen, Faculty of Production Engineering, 
28359 Bremen, Germany.\\
$^{5}$Leibniz Institute for Materials Engineering (IWT), 
28359 Bremen, Germany.\\
$^{6}$University of Bremen, MAPEX Center for Materials and Processes, 
28359 Bremen, Germany.
}

\subject{xxxxx, xxxxx, xxxx}


\corres{Marc Avila\\\email{marc.avila@zarm.uni-bremen.de}}

\begin{abstract}
Fluid flows between rotating concentric cylinders exhibit two distinct routes to turbulence. In flows dominated by inner-cylinder rotation, a sequence of linear instabilities leads to temporally chaotic dynamics as the rotation speed is increased. The resulting flow patterns occupy the whole system and sequentially lose spatial symmetry and coherence in the  transition process. In flows dominated by outer-cylinder rotation, the transition is abrupt and leads directly to turbulent flow regions that compete with laminar ones. We here review the main features of these two routes to turbulence. Bifurcation theory rationalises the origin of temporal chaos in both cases. However, the catastrophic transition of flows dominated by outer-cylinder rotation can only be understood by accounting for the spatial proliferation of turbulent regions with a statistical approach. We stress the role of \mod{the rotation number (the ratio of Coriolis to inertial forces)} and show that it determines the lower border for the existence of intermittent laminar-turbulent patterns.
\end{abstract}

\begin{fmtext}
\end{fmtext}

\maketitle

\section{Background}

Fluid flows tend to become turbulent as their speed increases. The emergence of turbulence is characterised by erratic fluctuations of fluid velocity and pressure, which enhance dissipation and thus lead to increased energy losses. In his seminal 1923 paper, Taylor~\cite{Taylor1923} computed the instability of laminar flows between co-rotating cylinders mathematically rigorously by starting from the Navier--Stokes equations. His own experimental verification of this result established hydrodynamic stability theory as a powerful tool to make accurate predictions of flow instabilities. To understand how remarkable this accomplishment is, it is necessary to review the historical background \cite{Eckert2010}. Interest in the problem of transition was spurred by Osborne Reynolds' 1883 paper on turbulence transition in pipe flow \cite{Reynolds1883}. Reynolds showed experimentally that in pipes this transition is sudden and depends on the level of disturbances present in the experiment. He concluded that ``\emph{the condition might be one of instability for disturbance of a certain magnitude and stable for smaller disturbances}.’’ Despite his remark, most of the subsequent theoretical approaches to the problem of transition, such as those pioneered by Orr~\cite{Orr1907}, Sommerfeld~\cite{Sommerfeld1909} and others, focused on the evolution of infinitesimal disturbances and thus failed to predict transition in pipe flow and even in the mathematically more tractable case of plane shear flow~\cite{Eckert2010}.\par
This failure of linear hydrodynamic stability theory to explain Reynolds' experiments, and the insurmountable difficulties inherent to nonlinear theories, continued to trouble the scientific community for forty years. Then Taylor~\cite{Taylor1923} made the key observation that \emph{``the study of the fluid stability when the disturbances are not considered as infinitely small is extremely difficult. It seems more promising therefore to examine the stability of liquid contained between concentric rotating cylinders. If instability is found for infinitesimal disturbances in this case it will be possible to examine the matter experimentally’’}. Taylor's proposal was motivated by Mallock's experiments in which the inner cylinder was rotated while the outer cylinder was held fixed~\cite{Mallock1896}. Mallock observed that the flow was turbulent even at the lowest speeds that he considered. These observations would go on to inspire Lord Rayleigh’s derivation of a criterion for the stability of rotating inviscid fluids~\cite{Rayleigh1917}. Taylor extended Rayleigh's analysis to include viscous effects by applying Orr's framework to the Navier--Stokes equations in cylindrical coordinates, keeping curvature terms up to first order. His calculations of the onset of instability were in excellent quantitative agreement with his own laboratory experiments. He showed this for different curvatures, as determined by the radius ratio, $\rratio=\ri/\ro$, where $r_i$ and $r_o$ are the radii of the inner ($i$) and outer ($o$) cylinder, respectively. Furthermore, the corresponding axially periodic eigenfunctions, which take on the form of toroidal vortices stacked along the axial direction, were in good agreement with his flow visualisation experiments using dye injections. Taylor's analysis was able to predict the stability of the flow even in regimes where Rayleigh's theory~\cite{Rayleigh1917} had failed, such as when the cylinders are made to counter-rotate or in experiments where the inner cylinder is made to rotate with very low velocity. More generally, Taylor's paper was also instrumental in confirming the validity of the Navier--Stokes equations, supplemented with no-slip boundary conditions, beyond the laminar regime.\par
An intriguing result of Taylor’s 1923 analysis, supported by his experiments, was that laminar flows with a stationary inner cylinder are linearly stable. This conflicted with earlier observations by Couette~\cite{Couette1890a}, who observed that the torque exerted on the inner cylinder scaled linearly at low angular speeds of the outer cylinder, \omegao, but became erratic as \omegao was increased above a certain critical threshold, before stabilising but exhibiting a stronger (nonlinear) scaling at even higher \omegao. Couette’s results agreed with observations by Mallock~\cite{Mallock1896}, who also noted that if the system was run right below the threshold for transition and a small perturbation was introduced, the torque measurements became erratic for extended periods of time before the system returned to its initial quiescent state. This transient nature of turbulence is a hallmark of linearly stable flows (see, \eg, Avila \etal~\cite{Avila2023} for a recent review) and will be discussed later in \S~\ref{sec:subcritical}\ref{sec:transients}. \par
Similar experiments were carried out by Wendt~\cite{Wendt1933} at even higher Reynolds numbers. Wendt also studied the effect of the end-wall conditions in his apparatus (\eg, attaching the cap at the bottom of his test section  either to the inner or the outer cylinder) and showed that these could change the critical rotation rate for transition by as much as \SI{10}{\percent}. Although Taylor initially attributed the increased torque reported for flows with stationary inner cylinder by Couette, Mallock, and Wendt to end-wall effects caused by the short aspect ratio, $\aspratio=h/\gap\lesssim\num{20}$, of their experimental apparatus (here, $h$ is the cylinders’ height and $\gap=\ro-\ri$ is the gap width between them), he was later able to reproduce qualitatively their observations with a new, longer and more accurate apparatus~\cite{Taylor1936a}. Furthermore, Taylor showed that the transition could be triggered by purposely applying slight perturbations and that turbulence remained sustained well below the threshold for natural (unforced) transition. This bi-stability of laminar flow and turbulence at fixed Reynolds numbers, reported earlier by Reynolds for pipe flow~\cite{Reynolds1883}, is another hallmark of linearly stable flows (\eg,~\cite{barkley2016theoretical}). Taylor also studied the effect of varying the system geometry and extrapolated his data toward the narrow gap limit, $\rratio\to\num{1}$, to infer that the critical Reynolds number for plane Couette flow (PCF) should lie between \num{315} and \num{500}, which is in agreement with more recent experiments~\cite{Daviaud1992, Bottin1998, klotz2022phase}. In the 1950s, Schultz-Grunow~\cite{SchultzGrunow1959} impressively showed that if the apparatus was fabricated with stringent tolerances, laminar flow could be maintained up to Reynolds numbers exceeding \num{40000}, whereas if he purposely misaligned his apparatus or used cylinders that were not perfectly round, his system would transition at Reynolds numbers consistent with earlier studies~\cite{Couette1890a, Mallock1896, Wendt1933, Taylor1936a}. \par
The early work by Taylor and others made it clear that flows between concentric cylinders exhibit two distinct routes to turbulence. The differences between these  routes were pinpointed in a decade-long study published by Coles~\cite{Coles1965} in 1965. In the first route, characteristic of flows dominated by inner-cylinder rotation, transition occurs via a sequence of instabilities that increase the temporal complexity of the flow and decrease its spatial symmetry. Coles characterised this sequence using hot wire anemometry and flow visualisations. For the second route, characteristic of flows dominated by outer-cylinder rotation, Coles noted that transition to turbulence is akin to that reported by Reynolds for pipe flow~\cite{Reynolds1883}. It is controlled by the evolution of finite-amplitude perturbations, which can cause the sudden emergence of localised turbulent domains interspersed in a laminar background. In this paper, we review previous works on the transition to turbulence in \TCF (TCF) and analyse them from the perspective of nonlinear dynamical systems. The main concepts are illustrated with results from new direct numerical simulations (DNS).

\section{Parameter regimes of \TCF (TCF)}
\label{sec:parspace}

Traditionally, the different dynamical regimes of TCF have been characterised in terms of two control parameters: the inner ($\Rei=u_i\gap/\visc$) and the outer ($\Reo=u_o\gap/\visc$) cylinder Reynolds number. They are based on the azimuthal velocity of each cylinder ($u_{i/o}=r_{i/o}\Omega_{i/o}$), \mod{the width of the gap (\gap) between the cylinders}, and the kinematic viscosity (\visc) of \mod{the fluid filling the gap.}\par
\begin{figure}
\centering
\includegraphics[width=1.0\textwidth]{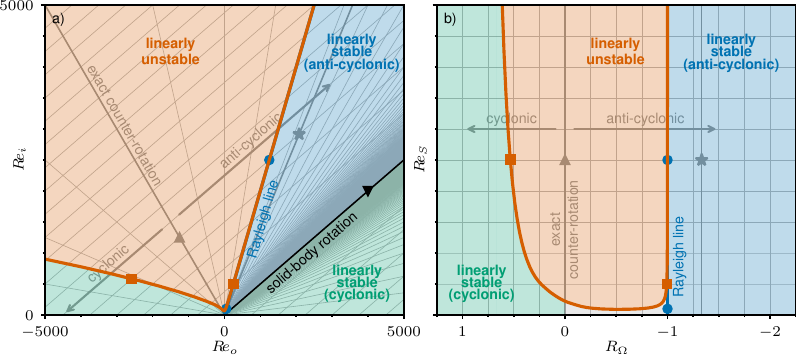}
\caption{Parameter space of \TCF. (a) In terms of the inner (\Rei) and outer (\Reo) cylinder Reynolds number; here for a system with one particular curvature (radius ratio $\rratio=\num{0.5}$). Above the neutral stability curve (red lines \mod{with squares}), the laminar flow is linearly (centrifugally) unstable. In the co-rotating regime, the Rayleigh line ($\Rei=\Reo/\rratio^2$, blue lines \mod{with circles}) is indistinguishable from the neutral stability curve (at this scale). For the specific case of solid-body rotation ($\Rei=\Reo/\eta$, black line \mod{with inverted triangle}), the flow has a constant angular velocity. \mod{Stable f}lows to the left of this line are anti-cyclonic, and cyclonic to the right. The specific case of exact counter-rotation (grey line with upright triangle) also separates the cyclonic from the anti-cyclonic regime. (b) Same as in (a) but in terms of the shear Reynolds number (\ReS) and the rotation number (\ROmega). Here, exact \mod{counter-rotation (zero mean rotation)} corresponds to $\ROmega=0$, the Rayleigh line is given by $\ROmega=\mod{\num{-1}}$ \mod{(in blue)}, \mod{Keplerian flow is best approximated with $\ROmega=-4/3$ (grey line with star),} and solid-body rotation is approached for $\ROmega\rightarrow\pm\infty$, when coming from the cyclonic (+) or the anti-cyclonic side (-), respectively.}
\label{fig:regimeMap}
\end{figure}

An example of a stability diagram using this traditional set of parameters is shown in Fig.~\ref{fig:regimeMap}(a) for a system with $\eta=\num{0.5}$.
The laminar base state of the system, \ie circular Couette flow (CCF), is linearly (centrifugally) unstable above the neutral stability curve. For a given value of \Reo, there is a critical \Rei above which the flow becomes unstable and secondary flows with increasing complexity emerge. The neutral stability curve we show here is analogous to those computed by Taylor~\cite{Taylor1923}, but we discretised the full Navier--Stokes equations linearised about the CCF state with a Galerkin method~\cite{Maretzke2014}. Transition to turbulence in the linearly unstable region (red in Fig.~\ref{fig:regimeMap}) is discussed in \S~\ref{sec:supercritical}. \par
Below the neutral stability curve, circular Couette flows are linearly stable and only finite-amplitude disturbances can trigger turbulence. Linearly stable CCF are classified into two distinct regimes with qualitatively different physics. In the linearly stable cyclonic regime (green in Fig.~\ref{fig:regimeMap}), the vorticity of the flow and the mean angular velocity are aligned. Here, the flow may exhibit a subcritical transition to turbulence. This transition scenario is reviewed in \S~\ref{sec:subcritical}. \mod{The linearly stable anti-cyclonic or quasi-Keplerian (QK) regime (blue in Fig.~\ref{fig:regimeMap}) is characterised by radially increasing angular momentum, but radially decreasing angular velocity. As reviewed in \S~\ref{sec:Kep}, the weight of evidence suggests that in this regime no transition to turbulence occurs even at Reynolds numbers exceeding a million.} \par
In 2005, Dubrulle \etal~\cite{Dubrulle2005} introduced an insightful alternative parametrisation of \TCF, which has improved our understanding of the physics of the system and embeds it in a much wider context of general rotating shear flows. \mod{For example, their parametrisation offers a simple relation to situate typical astrophysical (Keplerian) flows in the parameter space of \TCF. Essentially, they defined a shear Reynolds number (\ReS) and a rotation number (\ROmega) in such a way, that they translate to the traditional set of TCF parameters as}
\begin{align*}
\ReS = \frac{2\,\lvert\rratio\Reo-\Rei\rvert}{1+\rratio}
\quad\text{and}\quad
\ROmega = \frac{(1-\rratio)(\Rei+\Reo)}{\rratio\Reo-\Rei}
\text{,}
\end{align*}
respectively. \mod{We stress, that now the} ratio of \mod{inertial} to viscous forces and the ratio of \mod{Coriolis} to \mod{inertial forces} \mod{can be independently measured/controlled via \ReS and \ROmega, respectively.} \par
Figure~\ref{fig:regimeMap}(b) shows the same diagram as discussed before, but in terms of \ROmega and \ReS. This representation has the advantage that some of the important regime boundaries (except for the neutral stability curve) remain unchanged as \rratio is varied. Crucially, the line corresponding to exact counter-rotation ($\ROmega=0$) separates cyclonic ($\ROmega>0$) from anti-cyclonic ($\ROmega<0$) flows and smoothly connects \TCF to rotating plane Couette flow (RPCF) in the limit of vanishing curvature ($\rratio\rightarrow\num{1}$). Hence, RPCF is just a specific (limiting) configuration of TCF~\cite{Faisst2000}. Note that $\ReS=4\Reynolds$, where \Reynolds is the usual Reynolds number for PCF, which is defined with the wall velocity and half the gap width. The stability boundary derived by Rayleigh~\cite{Rayleigh1917} for inviscid fluids (\ie the Rayleigh line) is at $\ROmega=\num{-1}$. Solid-body rotation can be approached from the cyclonic side ($\ROmega\rightarrow\infty$) or from the anti-cyclonic side ($\ROmega\rightarrow-\infty$). In \mod{$(\Rei,\Reo)$}-space, however, these curves all depend on \rratio and thus move around as the curvature of the system is varied. This makes the comparison of results obtained for different geometries more challenging. \par
Before discussing transition in the different regions, it is important to point out that all the curves shown in Fig.~\ref{fig:regimeMap}, as well as their stability properties, were defined (or computed) under the assumption of an ideal (pure rotary) circular Couette flow as base state. In laboratory experiments, however, perfect CCF cannot be attained because of the axial end-walls, that confine the flow in axial direction and spawn secondary flows. Hence, \aspratio and the specific end-wall conditions used in laboratory experiments are additional important variables that need to be considered when comparing results from different Taylor--Couette simulations and experiments.

\section{Supercritical route to turbulence}
\label{sec:supercritical}

\TCF is least stable when the outer cylinder is held stationary ($\Reo=0$), as noted early on by Mallock~\cite{Mallock1896}. \mod{He reported turbulent flow for all rotation speeds of the inner cylinder that he could achieve in his experiments. In fact, the minimum realisable rotation speed in his apparatus already resulted in a supercritical Reynolds number (as shown later by Taylor \cite{Taylor1923}). Flows with stationary inner cylinder have} attracted a lot of interest~\cite{Tagg1994} because it simplifies the experimental setup considerably. In this section, we review the transition to turbulence for $\rratio=\num{0.875}$ and $\Reo=0$ as \Rei is increased from \num{100} to \num{4000}. Our choice is motivated by the influential experimental investigation of Gollub \& Swinney~\cite{gollub1975onset}, who measured time series of the radial velocity ($u_r$) at mid gap and revealed three distinct instabilities, each adding a new frequency to the Fourier spectrum of $u_r$ as \Rei was increased. In a further transition, the flow was shown to exhibit a continuous spectrum characteristic of chaotic dynamics, in apparent agreement with a theorem by Ruelle \& Takens~\cite{ruelle1971nature} regarding the existence of chaotic dynamics near three-tori in dynamical systems. Here, we present new computational results to illustrate the transition process. \par
To this end, periodic boundary conditions (BC) are employed in the axial ($z$) and azimuthal ($\theta$) direction and simulations of the incompressible Navier--Stokes equations carried out using our pseudo-spectral DNS code \nsc~\cite{Lopez2020}. The highest friction Reynolds number ($\ReTau=u_\tau d/\nu$) measured in all DNS presented here amounts to \num{95.7}, where $u_\tau$ is the friction velocity at the inner cylinder wall. \mod{We used a spatial resolution (in terms of viscous wall-units based on \ReTau and denoted by ${}^{+}$)} of at least $\num{0.02}\le\Delta r^+\le\num{1.3}$, $\Delta\theta r_i^+=\num{3.8}$ and $\Delta z^+=\num{3.5}$, which is the state of the art in DNS of wall-bounded turbulence~\cite{Ostilla-Monico2016, Feldmann2021}. In \nsc, the size of the time step is dynamically adapted during run time to ensure numerical stability and lies between \num{5e-7} and \num{5e-6} viscous time units ($d^2/\nu$) for all DNS presented here (depending on the exact setup). Except for the spiral turbulence simulation presented in \S~\ref{sec:subcritical}\ref{sec:spiralTurb}, all DNS were run for at least $\num{20}\,d^2/\nu$, before data were recorded and analysed.

\subsection{Bifurcation cascade to temporal chaos}
\label{sec:bifcas}

\begin{figure}
\centering
\includegraphics[width=1.0\textwidth]{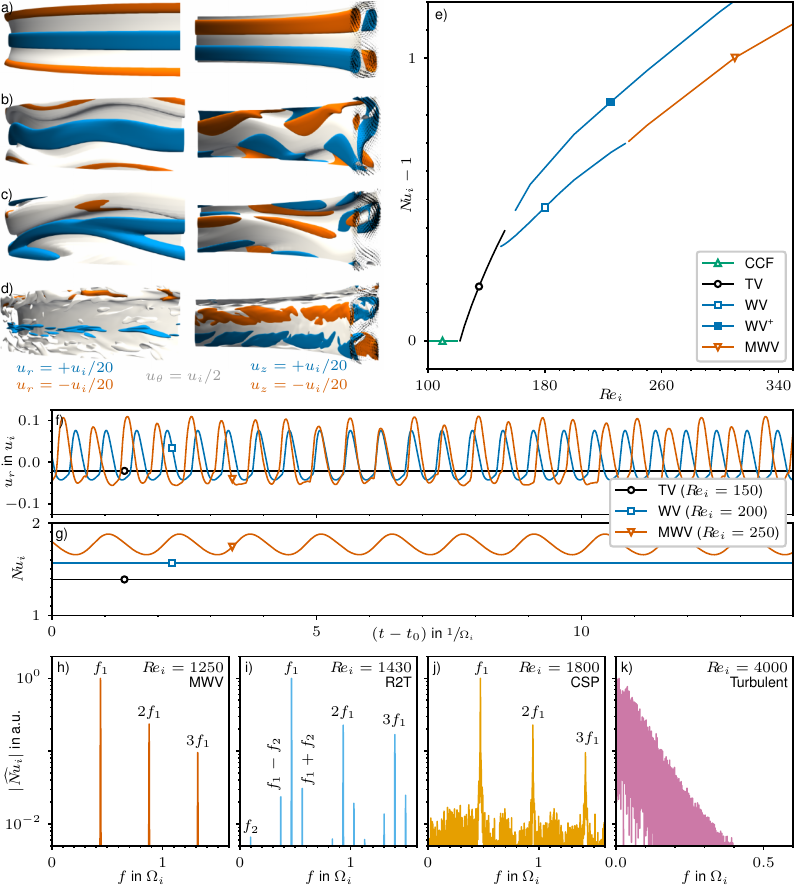}
\caption{Route to turbulence in \TCF with stationary outer cylinder ($\Reo=0$) as the inner cylinder Reynolds number (\Rei) is increased from \num{100} to \num{4000} in our simulations for fixed $\rratio=\frac{7}{8}$, $\aspratio=\frac{5}{2}$, and $L_\theta=\frac{\pi}{2}$. (a)--(d) Visualisations of colour-coded iso-contours of the radial ($u_r$), azimuthal ($u_\theta$), and axial ($u_z$) velocity components, show that the spatial structure of the flow field becomes progressively more complex as \Rei increases from (a) $\Rei=\num{150}$ (Taylor vortex, TV) to (b) $\Rei=\num{200}$ (wavy vortex, WV) to (c) $\Rei=\num{250}$ (modulated wavy vortex, MWV), and finally to (d) $\Rei=\num{4000}$ (turbulent flow). (e) Nusselt number of the inner cylinder (\Nui) as a function of \Rei for the different flow states of increasing complexity starting with circular Couette flow (CCF). (f)--(g) Time series of $u_r$ (recorded at mid gap) and \Nui for selected states. (h)--(k) Power spectra of the time series of \Nui for selected states including a relative two-torus (R2T) and a chaotic state \mod{exhibiting a continuous spectrum with peaks} (CSP).}
\label{fig:timeSeries}
\end{figure}

We begin by examining the onset of temporally chaotic dynamics in a small computational domain with axial aspect ratio $\aspratio=\frac{5}{2}$ and only one quarter of the circumference in the azimuthal direction ($\theta\in[0,L_\theta]$ with $L_\theta=\frac{\pi}{2}$). This choice is made to match the sequence of patterns reported by Gollub \& Swinney \cite{gollub1975onset} for a system with $\rratio=\frac{7}{8}$ and $\aspratio\approx\num{20}$. In their experiments, the primary instability took the form of Taylor vortices (TV) with an axial wavelength of approximately $\frac{5}{2}d$, corresponding to eight pairs of vortices stacked upon each other in their apparatus. Our simulations \mod{with $\aspratio=\frac{5}{2}$ are expected to capture the dynamics of two counter-rotating Taylor vortices} around the central section of their experiments, where the flow patterns are approximately periodic in $z$~\cite{gollub1975onset}. A summary of the flow patterns, dynamics and regimes as \Rei is increased in our simulations is presented in Fig.~\ref{fig:timeSeries} and detailed below.\par
At $\Rei\approx\num{122}$ the circular Couette flow becomes unstable and is replaced by one pair of axisymmetric TV, visualised in Fig~\ref{fig:timeSeries}(a) for the example of $\Rei=150$. In this transition, the invariance of CCF to axial translations is broken, while its rotational symmetry and its axial reflection symmetry (about the vertical mid plane) are preserved. The TV state is stationary, as can be seen in the time series of $u_r$ recorded at mid-gap position (Fig.~\ref{fig:timeSeries}(f)) and in the time series of the inner-cylinder Nusselt number (\Nui) as well (Fig.~\ref{fig:timeSeries}(g)). The latter represents the torque that is necessary to rotate the inner cylinder for a given flow state, normalised by the torque that would be required for CCF at the same \Rei. \par
Flow transitions, and more generally turbulence and transition, can be analysed from a dynamical systems perspective~\cite{Hopf1948}. In this framework, the Navier--Stokes equations span an infinite-dimensional phase space consisting of all admissible flow fields (\ie satisfying mass and momentum conservation). Here, time-dependent flows correspond to trajectories, whereas time-independent flows, like the TV flow, correspond to steady states or fixed-points in that phase space. The emergence of a new stable fixed-point (TV), which bifurcates from an existing one (CCF) that loses its stability, corresponds to a qualitative change in phase space or bifurcation. The bifurcation diagram in Fig.~\ref{fig:timeSeries}(e) captures this transition. Note that if TV is shifted axially by an arbitrary amount, the shifted TV pattern is still a solution to the system, so this bifurcation is known as a \emph{pitchfork of revolution}. For a more detailed treatment, the reader is referred to Kuznetsov~\cite{Kuz04} for an introduction to bifurcation theory, to Chossat \& Lauterbach~\cite{ChLa00} for a  discussion of equivariant bifurcation theory, to Chossat \& Iooss~\cite{ChIo94} for its application to \TCF, and to Crawford \& Knobloch~\cite{CrKn91} for a review of symmetry-breaking bifurcations in fluid dynamics. \par
As the inner cylinder is further accelerated beyond $\Rei\approx\num{150}$, the TV state turns unstable and the flow field becomes three-dimensional and time-dependent~\cite{Coles1965}. As shown in Fig.~\ref{fig:timeSeries}(b), the vortices acquire waviness in $\theta$, breaking the rotational and the axial reflection symmetry. However, if the pattern is rotated by $\frac{\pi}{2}$ and reflected about the mid plane, it remains unchanged~\cite{marcus1984simulation, Wereley1998}. The time series of $u_r$ for this wavy vortex (WV) state is now time-periodic (Fig.~\ref{fig:timeSeries}(f)). Generically, time-periodic states emerge from the loss of stability of steady states via Hopf bifurcations~\cite{hopf1942abzweigung, Kuz04}. However, WV states are not truly periodic solutions, but rather rotating waves~\cite{Ran82, King1984}. They rotate in $\theta$ like a solid body, and as such the time-dependence disappears when integral measures such as the torque (or \Nui) are considered (Fig.~\ref{fig:timeSeries}(g)). This is the reason why rotating waves are also known as relative equilibria~\cite{Kru90}. \mod{In this specific example, the Hopf bifurcation is subcritical and there is hysteresis. If the flow is initialised with a WV solution at $\Rei>\num{150}$ and then \Rei is reduced, WV remains stable down to $\Rei\approx\num{140}$. Hence in the interval $140\lesssim\Rei\lesssim150$ TV and WV co-exist bi-stably and both can be realised depending on the initial conditions. In addition, we found another wavy vortex solution (WV\textsuperscript{+}) in the same parameter range, which is stable for $\Rei\gtrsim\num{160}$ and is characterised by a higher Nusselt number (Fig.~\ref{fig:timeSeries}(e)). Overall, these calculations exemplify the multiplicity of states common to the (supercritical) route to turbulence in TCF, which is well-known since the seminal work of Coles~\cite{Coles1965}.}\par
Further increasing the Reynolds number beyond $\Rei\approx\num{237}$\mod{, the WV state bifurcates} to a modulated wavy vortex (MWV) state~\cite{Ran82, Andereck1986}. Now, $u_r$ becomes quasi-periodic (Fig.~\ref{fig:timeSeries}(f)) and exhibits two independent frequencies, whereas \Nui begins to oscillate periodically with only one frequency ($f_1$), as is clear from the time series shown in Fig.~\ref{fig:timeSeries}(g) and the corresponding spectrum shown in Fig.~\ref{fig:timeSeries}(h). MWV states are periodic orbits when observed from a reference frame moving at the speed of the underlying rotating wave~\cite{Ran82} and hence are often referred to as relative periodic orbits~\cite{viswanath2007recurrent}. At much larger Reynolds numbers ($\Rei\approx\num{1429.69}$), a new independent frequency ($f_2$) becomes apparent in the spectrum of \Nui (Fig.~\ref{fig:timeSeries}(i)), and the flow can be identified as a relative two-torus (R2T)~\cite{WLM01}; \ie a state with two independent frequencies ($f_1$ and $f_2$) when viewed in a co-moving frame. As \Rei is slightly increased (beyond $\Rei\approx\num{1437.50}$), the R2T state breaks down into chaos, as exemplified in the (turbulent) broadband spectrum for $\Rei=\num{4000}$ shown in Fig.~\ref{fig:timeSeries}(k). We found other solutions in the parameter range studied here. Notably, a chaotic solution with a clearly dominant frequency ($f_1$) emerges (Fig.~\ref{fig:timeSeries}(j)), which is stable for a wide range of \Rei. The computed flow states illustrate the dynamical richness of the Taylor--Couette system, and indeed many other flow states may exist. Exhausting the complete set of possible solutions (and the connections between them) is beyond the scope of this review article even for the small computational domain examined in this subsection. \par
The transition scenario observed in our simulations, including the Reynolds numbers of the instabilities, is in agreement with the experiments of Gollub \& Swinney~\cite{gollub1975onset}. They stressed the remarkable resemblance of their experimental observations with the theory of Ruelle \& Takens~\cite{ruelle1971nature}, who proved the existence of chaos close to three-frequency tori. However, as shown in Fig.~\ref{fig:timeSeries}(i), the R2T state is a two-frequency tori in a co-moving reference frame. Hence, their reported breakdown is actually in better agreement with a later paper by Newhouse, Ruelle \& Takens~\cite{newhouse1978occurrence}, who proved that generic breakdown into chaos is also possible from two-tori. The resulting temporally chaotic flow initially retains spatial coherence in $\theta$ and $z$, but this is gradually lost as \Rei is further increased~\cite{BrSw87, Lathrop1992}. Nevertheless, large-scale structures called turbulent Taylor vortices (because of their similarity to TV) are observed to persist in the time-averages of turbulent flows up to the highest Reynolds numbers ($\ReS=\orderof{e6}$) investigated so far (\eg~\cite{huisman2014multiple}). The focus of this paper is on the transitional regime and the reader is referred to Grossmann \etal~\cite{Grossmann2016} for a review on turbulent \TCF \mod{and to the recent work of Jeganathan \etal~\cite{Jeganathan2023} in this centennial issue, who explore the role of Coriolis forces on the formation of turbulent Taylor rolls}.

\subsection{Flows with defects}
\label{sec:stchaos}

The results presented in the previous section were obtained using a small computational domain containing exactly one pair of TV and one azimuthal wavelength of the WV state. In the experiments~\cite{gollub1975onset}, the WV repeated four times in $\theta$ and had eight pairs of vortices in $z$. This raises the question of whether the dynamics obtained in our small cell are also reproduced when spatially extended cells are considered. We carried out two DNS in a domain with $\aspratio=25$ and spanning the whole circumference. The simulations were initialised with the corresponding CCF state disturbed with a small perturbation repeating ten times in $z$ and four times in $\theta$, and numerical noise in the remaining modes. At $\Rei=\num{200}$, the CCF state rapidly destabilised and ten pairs of TV formed, which subsequently developed into a WV state of the exact same periodicity as in the simulations of the previous section. An examination of the time series of $u_r$ and \Nui confirmed that indeed the computed flow state is identical in both cases. \par
\begin{figure}
\centering
\includegraphics[width=1.0\textwidth]{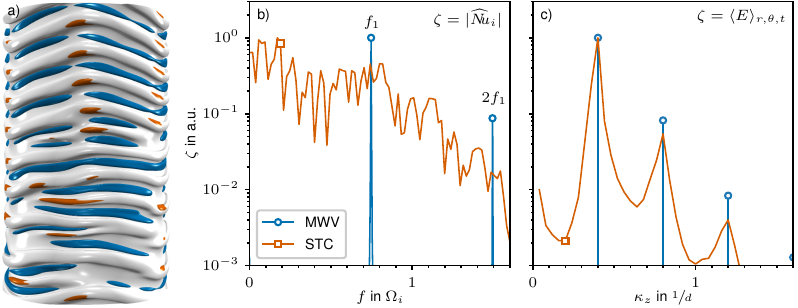}
\caption{Onset of spatio-temporal \mod{chaos (STC)} in large domains ($\aspratio=\num{25}$, $L_\theta=\num{2}\pi$) obtained in our simulations for fixed $\rratio=\frac{7}{8}$, $\Rei=\num{250}$, and $\Reo=\num{0}$. (a) Visualisation of the flow state with colour-coded iso-surfaces of the velocity field (grey for $u_{\theta}=u_{i}/2$ and red/blue for $u_{r}=\pm u_{i}/20$). (b) Power spectra of the corresponding time series of \Nui compared to a modulated wavy vortex (MWV) flow at the same $\Rei=\num{250}$ but in a smaller domain with $\aspratio=\frac{5}{2}$, $L_\theta=\frac{\pi}{2}$; \ie the state shown in Fig.~\ref{fig:timeSeries}(c). (c) Modal kinetic energy, $\langle E \rangle_{r,\theta,t}(\kappa_z)$, where  $\kappa_z$ is the axial wavenumber.}
\label{fig:longDomain}
\end{figure}

At $\Rei=\num{250}$, these two transitions were also observed and then followed by the transition to an MVW state also exactly as in the previous section. However, after a very long time (\num{20} viscous times units), the axial and azimuthal periodicity were lost as spatial defects appeared. A snapshot of this flow state is shown in Fig.~\ref{fig:longDomain}(a).
The corresponding temporal spectrum (Fig.~\ref{fig:longDomain}(b)) is continuous but features a broad, noisy peak close to the first harmonic of the spectrum of the MVW in the small domain. Similarly, all the spatial (axial) modes are excited, but as shown in Fig.~\ref{fig:longDomain}(c), the spectrum exhibits a clearly dominant peak (and harmonics) corresponding to ten pairs of vortices. Detailed investigations of transitions to chaotic flows with spatial defects, often referred to as spatio-temporal chaos \mod{(STC)}~\cite{cross1994spatiotemporal}, are scarce and deserve more attention. One example was studied in magnetohydrodynamic TCF, where the transition to defects was shown to emerge via a symmetry-breaking sub-harmonic bifurcation of a rotating wave~\cite{guseva2015transition}. While the temporal spectrum shown here is much noisier than that reported by Gollub \& Swinney~\cite{gollub1975onset} for the MVW state, their spectrum also contains substantial noise. This may be due to the presence of defects in the flow, similar to those reported here, or to measurement errors, and cannot be clarified here. However, we point out that the end-walls confining the fluid in $z$ result in distinctly different bifurcations, as demonstrated \eg in detailed experiments~\cite{abshagen2012localized} and in DNS~\cite{lopez2015dynamics} for systems with moderate aspect ratios ($\aspratio\sim\num{10}$). The effect of axial end-walls is reviewed below in \S~\ref{sec:supercritical}\ref{sec:endwalls}.

\subsection{Flow patterns in the linearly unstable regime}
\label{sec:linStabPatterns}

The results discussed in the previous sections are specific to flows with a stationary outer cylinder ($\Reo=\num{0}$) and that particular curvature ($\rratio=7/8$). Even for the small domain investigated in \S~\ref{sec:supercritical}\ref{sec:bifcas}, we found co-existence of many different flow states and alternative routes to chaos (not shown in detail here). This complexity is due to the nonlinear nature of the Navier--Stokes equations, resulting in a strong dependence on the initial conditions and in typical experiments on the path in $(\Rei,\Reo)$-space taken to prepare the system. This was known to Coles, who in his 1965 paper~\cite{Coles1965}, reported that for a given flow configuration (\ie particular combination of \Rei and \Reo) there could be as many as \num{26} different possible stable flow states (characterised by their wavenumbers $\kappa_\theta$ and $\kappa_z$) with complicated dynamical connections between them. \par
When the cylinders are allowed to rotate independently in the linearly unstable region, extremely rich dynamics are found. This was beautifully illustrated by Andereck \etal~\cite{Andereck1986}, who extensively mapped out the two-dimensional parameter space spanned by $(\Rei,\Reo)$ and revealed a large variety of flow regimes, characterised by their spatio-temporal symmetries and temporal frequencies as well as their azimuthal and axial wavenumbers. For example, in the counter-rotating regime, there exists an \Reo, at which the primary bifurcation to TV is replaced by a Hopf bifurcation to non-axisymmetric spiral vortices, which rotate in $\theta$ and travel in $z$. This spiral flow pattern was already observed by Taylor~\cite{Taylor1923} in his seminal paper and was later studied in detail theoretically~\cite{KGD66, LTKSG88} and experimentally~\cite{Sny68a}. As the speed of the inner cylinder (\Rei) is further increased, interpenetrating spiral patterns emerge\mod{, which consist of the nonlinear superposition of spiral vortices of opposite helicity}. Later the flow transitions to a chaotic state characterised by episodic bursts of turbulence superimposed on the laminar interpenetrating spirals~\cite{Andereck1986, Coughlin1996, meseguer2009instability, avila2013high}. This transition is just one of the multiple ones reported by Andereck~\cite{Andereck1986} and many others thereafter~\cite{Tagg1994}. The study of the various transitions between flow states has played a central role in the development of pattern formation theory \cite{Cross1993, Cross2009}.

\subsection{Effect of axial end-walls}
\label{sec:endwalls}

Whether comparing experimental results to the infinitely tall cylinders of theory, or to the periodic BC commonly used in simulations, the vertical boundaries of TCF are a widespread source of limitations, but also of new phenomena. Mitigating their effect started early, \eg, with Mallock~\cite{Mallock1889} using a pool of mercury at the bottom of his cylinders as a buffer \mod{and Couette introducing the use of guard rings~\cite{Couette1888a}}. Accordingly, the results discussed so far were for experiments with relatively tall cylinders ($\aspratio\gtrsim\num{20}$) or simulations with periodic BC. Cole~\cite{Col76} systematically investigated the effect of varying \aspratio and observed that the critical Reynolds number for TV is in good agreement with theoretical predictions for infinitely long cylinders even for aspect ratios as low as $\aspratio=\num{8}$. However, he also found that $\aspratio>\num{40}$ was necessary to obtain quantitative agreement with axially periodic simulations regarding the rotation speed of the WV. \par
From a theoretical stand point, the presence of end-walls destroys the translation invariance of the system. In addition, the necessary discontinuity of the BC between end-walls and inner/outer cylinder generates Ekman vortices~\cite{Burin2006}. As \Rei increases, the cellular TV pattern first appears near the end-walls and then rapidly fills the entire domain, as soon as the system reaches the critical \Rei predicted for the infinite-cylinder case~\cite{Col76, Ben78}. As a consequence, the transition to TV is not the result of a discrete bifurcation, but is a continuous process~\cite{Ben78}. \par
In systems with a stationary outer cylinder ($\Reo=\num{0}$) and the end-walls also at rest, the end-wall boundary layers tend to flow radially inwards. However, Benjamin~\cite{Ben78} found that cellular flows at low \aspratio could also have an outflow in one or both end-wall boundary layers. These flows, termed anomalous modes, are disconnected in phase space from the CCF solution and appear only at higher \Rei. For $\aspratio\approx\num{1}$, the competition between the normal two-cell mode and the one-cell anomalous mode has been extensively investigated~\cite{BeMu81, PSCM88, MTT02}, along with the transition to chaos~\cite{PBE92, MaLo06}. This specific case is particularly appealing for theoretical studies because the strong confinement drastically reduces the number of possible solutions to the Navier--Stokes equations. As \aspratio increases, the temporal dynamics become increasingly complex and a multitude of local~\cite{abshagen2012localized, lopez2015dynamics} and even global~\cite{abshagen2005symmetry} bifurcations have been reported. \par
End-wall effects have been also investigated in the counter-rotating regime, albeit less intensively. Non-axisymmetric spiral vortices can appear directly at the end-walls and propagate towards the centre of the system below the critical Reynolds number predicted by the infinite-cylinder approximation~\cite{heise2008localized}. The bifurcation scenario was found to depend strongly on the end-wall rotation speed~\cite{heise2009spirals}. Interpenetrating spirals have been also observed in recent experiments \cite{crowley2020novel} and simulations \cite{czarny2002spiral} for $\aspratio\approx\num{5}$, although this small aspect ratio results in a strongly modified laminar flow and a transition scenario that is qualitatively  different from that observed for tall cylinders~\cite{crowley2020novel}.

\section{Subcritical route to turbulence}
\label{sec:subcritical}

The transition to turbulence in flows dominated by outer-cylinder rotation is of a completely different nature. \mod{Here, the absence of a cascade of linear instabilities starting from laminar CCF greatly complicates the transition scenario and calls for a fully nonlinear approach. First, finite-amplitude perturbations are required to destabilise the laminar flow and the mechanisms sustaining turbulence are fully nonlinear~\cite{Boberg1988, Hamilton1995, Hall2010} and cannot be captured by a weakly nonlinear analysis. Second, the flow organises itself in complex laminar-turbulent patterns, that are spatio-temporally intermittent (Fig.~\ref{fig:spirals})}. As a consequence, understanding the subcritical transition in linearly stable TCF, and more generally in wall-bounded shear flows, requires stochastic approaches and additionally accounting for spatial effects. Neither of these two aspects is covered by the classical route to \mod{temporal} chaos discovered by Ruelle \& Takens \cite{ruelle1971nature, newhouse1978occurrence} and discussed in the previous section. The appropriate framework to describe \mod{the onset of laminar-turbulent patterns} was proposed by Kaneko~\cite{Kaneko:1985}, who extended the dynamical systems approach to linearly stable\mod{, spatially extended flows}. The first experimental studies embracing his spatio-temporal perspective were carried out in the nineties by Bottin \etal~\cite{Bottin1998, Bottin1998b} for PCF.

\subsection{Spiral turbulence and turbulent spots}
\label{sec:spiralTurb}

\begin{figure}
\centering
\includegraphics[width=1.00\textwidth]{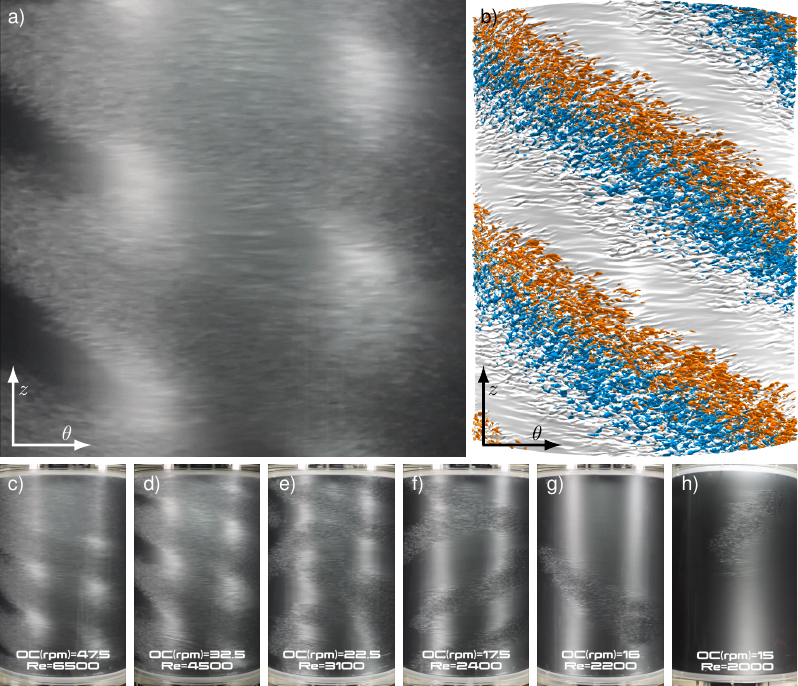}
\caption{Spiral turbulence and turbulent spots in \TCF with stationary inner cylinder ($\Rei=\num{0}$). (a) Spiral turbulence in experiments~\cite{Burin2012} for $\Reo=\num{4500}$, $\rratio=\num{0.97}$, and $\aspratio=\num{92}$. Kalliroscope flakes were added to the water for imaging. (b) Spiral turbulence in a direct numerical simulation for the same parameters as in (a) but for $\aspratio=\num{61.3}$, $L_\theta=\pi/2$ and using periodic boundary conditions (BC) in \mod{the azimuthal ($\theta$) and axial ($z$) direction}. Shown are colour-coded iso-contours of the velocity field ($u_\theta=u_o/2$ in grey and $u_r=\pm u_o/10$ in red/blue). (c)--(h) Sequence of flow patterns as \Reo is decreased in the experiments~\cite{Burin2012} from $\Reo=\num{6500}$ \mod{down} to \num{2000} (Supplementary video online).}
\label{fig:spirals}
\end{figure}

For flows with stationary inner cylinder ($\Rei=\num{0}$) and rapid outer cylinder rotation Mallock~\cite{Mallock1896} noted already in 1896 that ``\emph{when the velocity approached that at which instability was liable to occur, it was interesting to notice how small a disturbance of the system was sufficient to change the entire character of the motion. A slight blow on the support which carried the apparatus, or a retardation for a few moments of the rotation of the outer cylinder, was almost sure to produce the effect.}'' The flow patterns emerging from this type of transition consist of well-defined turbulent domains (spots) that evolve in an otherwise laminar background, very similar to what Reynolds~\cite{Reynolds1883} reported for pipe flow. Coles~\cite{Coles1965} observed that individual spots sometimes spontaneously laminarised, but could also grow in size and sometimes merge with other spots. He reported that under some conditions turbulent spots came together and formed a spiral band of turbulence that cork-screwed around the apparatus, see Fig.~\ref{fig:spirals}(a)--(d). The eminent physicist Richard Feynman, who was a Professor at Caltech at the same time as Coles, was intrigued by the origin of this surprisingly organised, but at the same time turbulent pattern, which he called ``\emph{barber-pole turbulence}''~\cite{Feyman1977}. \par 
The emergence and death of spiral turbulence is demonstrated in Fig.~\ref{fig:spirals}(c)--(h) with snapshots from the experiments of Burin \& Czarnocki~\cite{Burin2012} for $\Rei=\num{0}$ and $\rratio=\num{0.97}$.
\mod{They quasi-statically increased the rotation speed of the outer cylinder and reported an abrupt transition from laminar to spatio-temporally intermittent flow at around $\Reo\approx4500$.  Since no specific perturbations were applied to their experiments, this corresponds to the natural transition point at which ambient noise and/or imperfections sufficed to trigger transition in their setup. Further increasing \Reo to \num{8000} led to a fully turbulent flow depleted of laminar regions. Subsequently, they decreased \Reo to study the spatio-temporal dynamics of turbulence. The snapshots in Fig.~\ref{fig:spirals}(c)--(h), and the corresponding supplementary movie, illustrate how laminar regions gradually appear and grow in size as \Reo is reduced, leading to the formation of laminar-turbulent spiral patterns. This transition process from fully turbulent TCF to laminar-turbulent spiral patterns can be understood as a spatial modulation of the fully turbulent state \cite{prigent2002large}, arising from a linear instability of the latter \cite{Kashyap2022}.}
As \Reo was further decreased, the spiral turbulence retained its form, until it began to progressively disintegrate into spots. These spots were sometimes elongated, in line with the spiral angle, and their boundary appeared to propagate in this direction as well (see \eg, the video at $\Reo=\num{2100}$). Finally, the flow laminarised fully at $\Reo\approx\num{2000}$. \mod{We stress that the parameter regime $\Reo\in[\num{2000},\num{4500}]$ is characterised by strong hysteresis. Here, the history of the flow (\ie its path taken in parameter space) determines the observed flow state for a given \Reo. If the flow is initially laminar and ``undisturbed'', it remains laminar forever. By contrast, if it is initially turbulent (or laminar, but sufficiently disturbed), it remains turbulent.}\par
\begin{figure}
\centering
\includegraphics[width=1.0\textwidth]{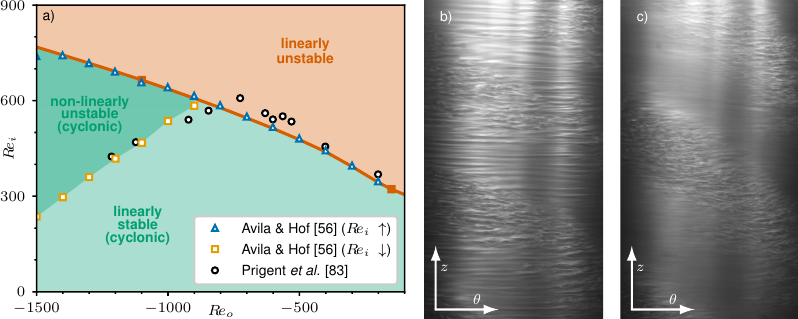}
\caption{\mod{Sub- and supercritical spiral turbulence in \TCF experiments~\cite{Prigent2003, avila2013high}, both with counter-rotating cylinders and narrow gaps ($\rratio\approx\num{0.98}$). (a) Laminar flow can exist everywhere below the theoretical neutral stability curve (red line with full squares as in Fig.~\ref{fig:regimeMap}). In the linearly stable cyclonic regime (light green) it is the only flow state. In the non-linearly unstable hysteretic regime (dark green) laminar flow or subcritical spiral turbulence can be observed, depending on the flow history and the existence of finite amplitude perturbations. Above the neutral stability curve, laminar interpenetrating spirals arise, sometimes coexisting with turbulent spots/spirals. (b) Supercritical spiral turbulence featuring intermittent turbulent spirals on top of laminar interpenetrating spirals \cite{avila2020second}. (c) Subcritical spiral turbulence with the typical coexisting laminar and turbulent domains \cite{avila2020second}.}}
\label{fig:spiralsCounter}
\end{figure}

\mod{This behaviour is not specific to flows with a stationary inner cylinder, but extends to most of the counter-rotating regime of \TCF~\cite{Andereck1986, meseguer2009instability, avila2013high}. Figure~\ref{fig:spiralsCounter}(a) shows the regime diagram for $\rratio=\num{0.98}$.}
\mod{We first focus on the transition scenario for $\Reo=\num{-1000}$, which was the focus of an experimental study by Avila \& Hof~\cite{avila2013high}. As \Rei exceeds \num{650}, the laminar flow becomes unstable to a spiral mode and accordingly a transition to interpenetrating laminar spiral patterns was observed in the experiments. Upon a slight increase in \Rei turbulent spots spontaneously formed and decayed in an unpredictable manner, while the interpenetrating spirals remained largely unaffected in the background, as exemplarily shown Fig.~\ref{fig:spiralsCounter}(b). This intriguing flow state was discussed briefly in \S~\ref{sec:supercritical}\ref{sec:linStabPatterns} and has been reported by many authors~\cite{Andereck1986, Coughlin1996, meseguer2009instability, avila2013high}. Avila \& Hof~\cite{avila2013high} then reduced \Rei and noted that the interpenetrating spirals vanished, while the turbulent spots organised themselves into turbulent spiral arms interspersed in the remaining laminar background (Fig.~\ref{fig:spiralsCounter}(c)). The resulting spiral turbulence remained robust as \Rei was further reduced, until it disintegrated into spots and ultimately laminarised fully at $\Rei=525$. Similar results were obtained for other \Reo. As shown in Fig.~\ref{fig:spiralsCounter}(a), the faster the rotation of the outer cylinder, the larger the region of hysteresis in which spiral turbulence or fully laminar flow can be observed (depending on the initial conditions). For the same geometry, Prigent~\etal \cite{Prigent2003} did not detect hysteresis. In their experiments, the natural transition point coincides approximately with the laminarisation border of spiral turbulence. This suggests a high level of background disturbance in their setup and illustrates the dependence of the natural transition point on technical  details of the apparatus. Overall, the results of Avila \& Hof~\cite{avila2013high} are in qualitative agreement with earlier numerical simulations of Meseguer \etal \cite{meseguer2009instability}, who studied the counter-rotating regime for $\rratio=0.883$ in detail.}\par
\mod{We remark that in the experiments of Burin \& Czarnocki~\cite{Burin2012} and Avila \& Hof~\cite{avila2013high} the subcritical turbulence was created differently. In the latter, the laminar flow reproducibly became unstable as \Rei was increased above the linear instability threshold. By successively decreasing \Rei below this threshold, the subcritical turbulence was investigated. By contrast, in the experiments of Burin \& Czarnocki~\cite{Burin2012} with a stationary inner cylinder, a linear instability was unavailable and finite-amplitude disturbances (\eg, unavoidable imperfections in the setup, vibrations or ambient noise) trigger turbulence, as in Prigent's setup \cite{Prigent2003}. Despite this difference, the reverse transition from turbulent to laminar flow, as the Reynolds number decreases, is less sensitive to disturbances. Generally, in the subcritical regime, the properties of the laminar-turbulent patterns are intrinsic to the system and independent of how the turbulent flow is initialised.}\par
\begin{figure}
\centering
\includegraphics[width=1.0\textwidth]{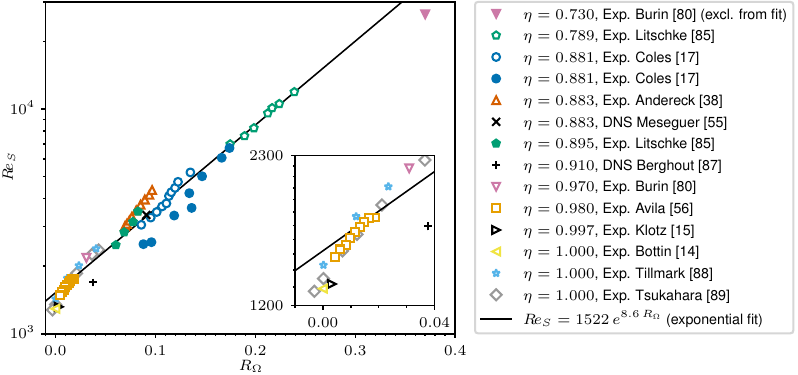}
\caption{Laminarisation boundary of turbulent spots in terms of the critical shear Reynolds number (\ReS) as a function of the rotation number (\ROmega). The data sets from many different experiments~\cite{Coles1965, Andereck1986, Tillmark1996, Litschke1998, Bottin1998, Tsukahara2010FlowRotation, Burin2012, avila2013high, klotz2022phase} and simulations~\cite{meseguer2009instability, berghout2020direct} span a wide range of curvatures ($\rratio\in[\num{0.73}, \num{1}]$) and Reynolds numbers ($\Rei\in[\num{-1000},\num{1200}]$, $\Reo\in[\num{-50000},\num{-1200}]$). Most measurements were taken in the subcritical counter-rotating regime or for stationary inner cylinder, but Coles~\cite{Coles1965} also performed measurements in the co-rotating cyclonic regime. \mod{Note that in typical experimental setups with fixed cylinder length, an increase in gap width (\ie, a decrease in \rratio) comes at the price of a decrease in \aspratio. As a consequence, the critical \ReS is more affected by the specific end-cap configuration. Thus, the data of Burin \& Czarnocki~\cite{Burin2012} for wide gaps ($\rratio=\num{0.73}$ and $\rratio=\num{0.55}$, the latter not shown here) were excluded from the exponential fit.}}
\label{fig:subcrit}
\end{figure}

The lower boundary for the laminarisation of turbulent spots has been measured by many authors experimentally~\cite{Coles1965, Andereck1986, Litschke1998, Bottin1998, meseguer2009instability, Burin2012, avila2013high, klotz2022phase} and in DNS~\cite{barkley2005computational, meseguer2009instability, berghout2020direct}. These measurements are summarised in Fig.~\ref{fig:subcrit}, where the lowest \ReS for turbulence survival is shown as a function of \ROmega.
Recall that $\ROmega=\num{0}$ corresponds to exact counter-rotation and the first data point (lowest $\ROmega$) here corresponds to PCF flow with $\ReS\approx\num{1300}$~\cite{Bottin1998}, which we calculated as $\ReS=4\Reynolds$, where \Reynolds is the Reynolds number commonly defined for plane Couette flow. This is followed by the experiments of Klotz \etal~\cite{klotz2022phase} for an extremely narrow gap ($\rratio=\num{0.997}$) and stationary inner cylinder with $\ReS\approx\num{1320}$ and $\ROmega=\num{0.003}$. As \ROmega is increased and cyclonic rotation is added, higher \ReS are required to sustain turbulence. As shown in Fig.~\ref{fig:subcrit}, experimental data from various studies collapse well into a single curve despite the wide ranges covered in curvature $\rratio\in[0.789,1]$ and in Reynolds numbers ($\Rei\in[\num{-1000},\num{1200}]$ and $\Reo\in[\num{-50000},\num{-1200}]$). One exception to this are the transition points obtained by Burin \& Czarnocki~\cite{Burin2012} for wide gaps, $\rratio=\num{0.73}$ and \num{0.55} (the latter not shown in the plot, as it corresponds to much larger \ROmega). For a fixed experimental apparatus, wider gaps result in lower \aspratio. In combination with the increased \ReS, which is required for subcritical transition, this leads to end-wall effects becoming very significant for the laminarisation boundary. In that case, the end-walls greatly perturb the laminar base state and may trigger subcritical transition~\cite{Burin2012} or even result in linear instabilities~\cite{AGLM08, Avila2012, lopez2016subcritical}. \mod{For channel flow, it has been recently shown that end-walls destabilise laminar-turbulent bands and can cause relaminarisation~\cite{Kohyama2022, Wu2022}.}
\par
The remarkable collapse of the data (Fig.~\ref{fig:subcrit}) illustrates the suitability of the parametrisation introduced by Dubrulle \etal~\cite{Dubrulle2005} and demonstrates that the specific case of stationary inner cylinder (\ie $\ROmega=\sfrac{1}{\rratio}-1$) does not exhibit some form of singularity, as Fig.~\ref{fig:regimeMap}(a) may initially suggest. This point is further confirmed by the experiments of Coles~\cite{Coles1965}, who continued the laminarisation curve from the counter-rotating to the co-rotating regime. Note also that curvature appears to influence the laminarisation boundary only indirectly through \ROmega, at least for $\rratio\gtrsim\num{0.789}$. \par
Extending measurements to larger \ROmega would be valuable, but is challenging. Specifically, as \ROmega increases, \ReS grows exponentially, necessarily resulting in a huge difference between the angular speeds of the inner and outer cylinders. This exacerbates end-wall effects in experiments and has extremely high computational cost in axially periodic DNS. Finally, we stress that the boundary at which turbulence can become sustained, depends only on the system parameters. By contrast, the natural (uncontrolled) transition boundary, where turbulence is observed to appear as \ReS is increased, depends on the level of disturbance present in the experiment and can vary by as much as one order of magnitude in \ReS for the same geometry and a fixed value of \ROmega~\cite{SchultzGrunow1959}.

\subsection{Transient turbulence: stochastic decay of turbulent spots}
\label{sec:transients}

The experiments reported by Mallock~\cite{Mallock1896}, Taylor~\cite{Taylor1936a} and Coles~\cite{Coles1965} demonstrated that, once triggered, turbulent episodes can last almost indefinitely or relaminarise after irregular intervals. This essentially implies that the lower boundary discussed in the previous section depends on the observation time and on the experimental realisation, \mod{which in part explains the scatter in Fig.~\ref{fig:subcrit}}. \par
\begin{figure}
\centering
\includegraphics[width=1.0\textwidth]{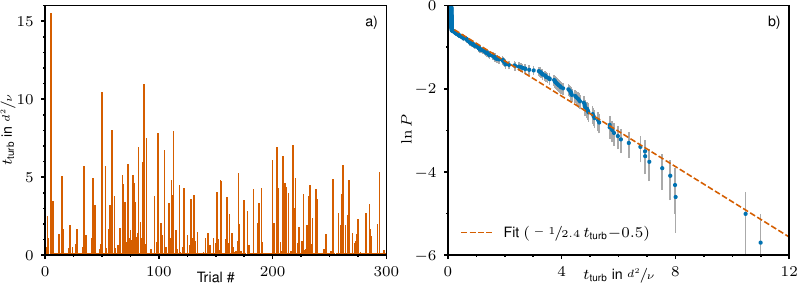}
\caption{Lifetime statistics of turbulent episodes in \TCF experiments~\cite{borrero2010transient} at $\Rei=0$, $\Reo=\num{8106}$, $\rratio=\num{0.871}$, and $\aspratio=\num{33.6}$. (a) Duration ($t_{\text{turb}}$) of turbulent episodes after triggering turbulence by briefly counter-rotating the inner cylinder in \num{300} experimental realisations. For some runs turbulence collapses almost immediately, but for others it persists up to $\orderof{e1}$ viscous time units ($d^2/\nu$), which for this Reynolds number corresponds to $\orderof{e5}$ convective time units ($d/u_o$). (b) Probability ($P$) that turbulence survives up to time $t_{\text{turb}}$. The runs where turbulence collapses immediately ($t_{\text{turb}}\approx\num{0}$), are highlighted by an almost vertical drop in $P$. For the rest of the runs, $P$ decays approximately exponentially, as indicated be the linear fit (broken line) with the slope $-1/\tau_d$. Here, $\lifetime\approx\num{2.4}$ is the average survival time. Note the semi-log scale of the plot. The error bars (grey) represent the sampling error.}
\label{fig:lifetimes}
\end{figure}

The first systematic study of the stochastic nature of turbulence decay in \TCF was performed in 2010 by Borrero-Echeverry \etal~\cite{borrero2010transient}. They investigated a system driven by pure outer cylinder rotation ($\Rei=\num{0}$) and repeated the experiments up to \num{1200} times for every \Reo in order to statistically characterise the duration of turbulent episodes. Their flow visualisations indicated that turbulence collapses suddenly and without warning. The duration (or lifetime) $t_{\text{turb}}$ of turbulence at $\Reo=\num{8106}$ in each experimental run is shown in Fig.~\ref{fig:lifetimes}(a).
The corresponding lifetime distribution, shown in Fig.~\ref{fig:lifetimes}(b), is approximately exponential. This implies that relaminarisation follows a memoryless process, exactly as reported earlier for plane Couette~\cite{Bottin1998} and pipe~\cite{Peixinho2006, Hof2006} flows. By collecting \mod{lifetime} distributions at various \Reo, Borrero-Echeverry \etal~\cite{borrero2010transient} showed that the associated time constant \lifetime (\ie the average lifetime) increases super-exponentially with \Reo, but remains finite as \Reo increases. This result, in qualitative agreement with previous studies of pipe flow~\cite{Hof2008, Avila2010}, was confirmed by Alidai \etal~\cite{alidai2016turbulent}, who additionally showed that \lifetime does not depend on the specific details of how the turbulence is initially triggered, but is significantly affected by imperfections in the experimental apparatus, and by the end-wall conditions. Finally, they also showed that as \aspratio is reduced, \lifetime decreases.
\par
Goldenfeld \etal~\cite{Goldenfeld2010} proposed that turbulent spots relaminarise when all turbulent fluctuations within the spot fall below a certain level. With this assumption they exploited extreme-value theory, where super-exponential scaling appears as a matter of course~\cite{Gumbel1958}, to develop a model consistent with the scaling of \lifetime observed in experiments~\cite{Hof2008, borrero2010transient}. This idea has recently been \mod{validated} and elaborated further in DNS studies~\cite{Nemoto2021, Gome2022}.

\subsection{Onset of sustained turbulence: spatio-temporal proliferation of spots}

Despite the transient nature of individual turbulent spots, Moxey \& Barkley~\cite{moxey2010distinct} showed a mechanism by which the proliferation of turbulent spots (\ie puffs in their case) can lead to sustained turbulence in pipe flow. Avila \etal~\cite{Avila2011} further showed that the probability of a puff to grow and shed a daughter one, statistically identical to its parent, increases as the Reynolds number increases. Specifically, the associated proliferation time constant decreases super-exponentially with Reynolds number. The competition between the stochastic processes of turbulence decay and turbulence proliferation determines the critical Reynolds number for the onset of sustained turbulence in pipe flow~\cite{Avila2011}. The same mechanism was later demonstrated in a narrow-gap Taylor--Couette system~\cite{shi2013}, in which the computational domain was spatially extended in one direction only (as happens naturally in pipe flow). \par
Beyond the critical point, turbulence proliferates faster than it decays and the fraction of the flow domain occupied by turbulence increases with Reynolds number. This was explicitly demonstrated by Lemoult \etal~\cite{lemoult2016directed}, who precisely measured the evolution of the turbulence fraction as a function of \Reo for a system with stationary inner cylinder and a very narrow gap ($\rratio=\num{0.998}$). The short aspect ratio they used ($\aspratio=\num{8}$) allowed turbulence to localise only in $\theta$, whilst filling the apparatus in $z$ completely. The turbulence fraction was found to increase continuously from zero at the critical point, corresponding to a second-order phase transition. The spatio-temporal dynamics of the turbulent spots, and of the laminar gaps between them, could be shown to fall into the universality class of directed percolation (DP), which is the simplest case of a non-equilibrium phase transition. We stress that in this system the dynamics of spots is exactly as in pipe flow due to the axial confinement. The critical exponents could be accurately determined with extremely long measurement times, because \TCF is a closed system in contrast to pipe flow, where turbulent puffs ultimately leave the pipe at its downstream end. \par
It is worth noting that the mechanisms of turbulent proliferation are qualitatively different in flows extended in one and in two directions. In the former, the turbulence fraction can only increase as the spots grow in azimuthal direction ($\theta$), whereas in the latter, spots can also extend in the axial direction ($z$). In both cases the transition is of second order~\cite{avila2020second}. Moreover, recent measurements by Klotz \etal~\cite{klotz2022phase} in very tall cylinders ($\aspratio=\num{750}$) with a very narrow gap ($\rratio=\num{0.997}$) exhibit also a DP transition; but with exponents characteristic of systems extended in two directions (here in $\theta$ and $z$). The interested reader is referred to two recent reviews about laminar-turbulent patterns in wall-bounded flows extended in two spatial directions~\cite{tuckerman2020patterns} and about the role of DP for turbulence transition in general~\cite{Hof2022}.

\subsection{Simple invariant solutions and the origin of transient turbulence}

The results discussed in the previous section resolve the question of the onset of sustained turbulence in linearly stable flows, but do not provide information as to how chaotic dynamics appears in the phase space of the system. In an influential paper, Nagata~\cite{nagata1990} showed in 1990 that a specific wavy vortex solution, similar to the WV shown in Fig.~\ref{fig:timeSeries}(b) for stationary outer cylinder, could be numerically continued from Taylor--Couette to plane Couette flow (without rotation). In this limit, Nagata's WV solution remains three-dimensional, but becomes steady. More importantly, it is disconnected from the laminar (plane) Couette flow in phase space, a result that holds already for TCF in very narrow gaps ($\rratio\gtrsim\num{0.993}$)~\cite{Faisst2000}. In other words, in PCF there is no sequence of instabilities leading from its laminar base state to WV flow, because the base state is linearly stable. Nagata's  discovery paved the way for applying the dynamical-systems approach to linearly stable flows. \par
\begin{figure}
\centering
\includegraphics[scale=1.0]{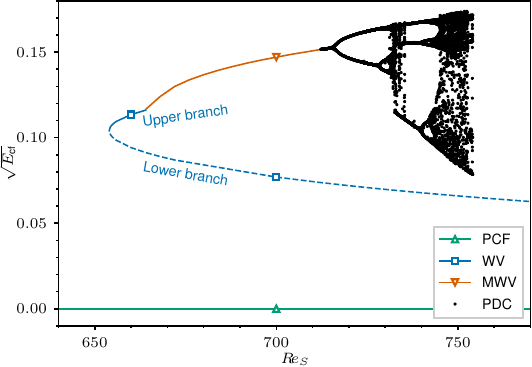}
\caption{Bifurcation diagram showing the onset of chaotic transients in plane Couette flow (PCF) in a small, symmetry-restricted cell. The kinetic energy contained in the cross-flow velocity components, $\sqrt{E_\text{cf}}$, is shown as a function of the shear Reynolds number (\ReS). A pair of steady wavy vortex (WV) solutions is born at a saddle-node bifurcation at $\ReS\approx\num{655}$. The upper branch is stable in the symmetry-restricted space, but loses stability in a Hopf bifurcation to a modulated wavy vortex (MWV) state at $\ReS\approx\num{664}$. Subsequently, a period-doubling cascade (PDC) generates a chaotic attractor, although there is a short window ($\num{735}<\ReS<\num{740}$) with a stable period three state. The chaotic attractor is destroyed at $\ReS\approx\num{754}$ and thereafter all trajectories laminarise. Note, that the shear Reynolds number is related to the usual plane Couette Reynolds number as $\ReS=4\Reynolds$. The data shown in this figure is re-plotted with permission from Kreilos \& Eckhardt~\cite{kreilos2012periodic}.}
\label{fig:kreilosReplot}
\end{figure}

At the lowest Reynolds number (called saddle-point) at which Nagata's WV solution exists, it is stable in a certain symmetry-restricted subspace (but unstable in full space). As shown in Fig.~\ref{fig:kreilosReplot} (data re-plotted with permission from Kreilos \& Eckhardt~\cite{kreilos2012periodic}), at Reynolds numbers slightly above the saddle-point this solution undergoes a Hopf bifurcation leading to a periodic solution~\cite{clever1997tertiary}.
This is exactly like the bifurcation from a WV state into a relative periodic solution (MWV) shown earlier in Fig.~\ref{fig:timeSeries}(e) for \TCF. As \ReS is further increased, a period-doubling cascade (PDC) renders the flow chaotic. Despite the difference in the precise bifurcations occurring here in PCF and in the case of TCF, the situation is similar in both cases. Chaos emerges from a simple solution (WV) after a short bifurcation sequence during which symmetries are lost and the temporal complexity increases. We however stress that in PCF, these changes in phase space occur far away from the laminar base state, which actually remains stable to infinitesimal disturbances for arbitrarily large Reynolds numbers~\cite{romanov1973stability}. \par
As \ReS is further increased, the chaotic attractor's volume grows in phase space, \ie larger fluctuations of the velocity occur. Finally, the attractor collides with a twin lower branch WV solution, born at the same saddle-node bifurcation, and is destroyed~\cite{lustro2019onset, Grebogi1983}. Beyond this \ReS the chaotic dynamics is not sustained, but transient, and the resulting lifetime distributions are exponential~\cite{kreilos2014increasing}. Although this is consistent with the experimental observations at larger \ReS described in the previous section, there are crucial qualitative differences. The DNS described in this section were performed by imposing artificial symmetries in small computational cells. This stabilises the solutions, simplifies the temporal dynamics and precludes the spatial localisation typical of turbulent spots. In conclusion, a wide gap between theory and experiment remains at this stage. The interested reader is referred to the recent review by Avila \etal~\cite{Avila2023} with main focus on pipe flow. \par
Several families of simple solutions with distinct symmetries have been computed in \TCF to date~\cite{meseguer2009families, deguchi2014subcritical, wang2022}. A particularly interesting solution computed for plane Couette flow~\cite{reetz2018invariant} is spatially localised in the form of a spiral, similar to the spiral turbulence observed in experiments and shown in Fig.~\ref{fig:spirals}. Generally, these solutions are all unstable, but it has been demonstrated that the phase space of the turbulent dynamics is organised around them and their stable and unstable manifolds \cite{kawahara2001periodic, gibson2008visualizing}; reviews are given by Graham \& Floryan~\cite{Graham2021} and Kawahara \etal~\cite{kawahara2012significance}. More recent studies using a Taylor--Couette system with $\aspratio = 1$ have provided experimental evidence supporting the importance of simple solutions in organising the turbulent dynamics \cite{krygier2021, Crowley2022a, Crowley2023}. \par
We conclude this section by noting that Deguchi~\cite{deguchi2017linear} recently made the surprising discovery of a linear instability of \TCF with stationary inner cylinder. However, for typical experimental gap  widths (\ie $\rratio\gtrsim\num{0.1}$), this instability only occurs for $\Reo\gtrsim\num{e7}$. Even if these extreme Reynolds numbers could be attained in experiments (not achieved so far), subcritical transition would certainly occur well before this instability could be observed.

\section{Quasi-Keplerian regime}
\label{sec:Kep}

\mod{Because of its potential relevance to astrophysical flows, the QK regime of \TCF ($-\infty<\ROmega<\num{-1}$, blue region in Fig.~\ref{fig:regimeMap}) has attracted particular interest in recent years. A thin disc of gas in orbital motion around a central gravitating body is a paradigmatic scenario in astronomy usually associated with the beginning or end of stellar lifetime ~\cite{Pringle1981, papaloizou1995theory, soward2005fluid}. In principle, accretion in astrophysical discs requires a radial transport of orbiting gas towards the central mass, and the corresponding inward loss of angular momentum is explained by an outward transport of the same. If the orbiting motion is laminar, the only mechanism available to maintain outward momentum transport is molecular viscosity, which is orders of magnitude too low for accretion to take place on the inferred time scales. Thus the transport is naturally turbulent, but the possible origin of this turbulence has generated considerable discussion.} \par
\mod{In astrophysical discs, the angular velocity of the orbiting gas is often assumed to scale as $\Omega\propto r^{-q}$, where $r$ is the distance to the central object and $q=-2/\ROmega$ fully characterises the flow with $q=3/2$ corresponding to Keplerian flow ($\ROmega=-4/3$)~\cite{Dubrulle2005}.
Significantly, the flow in the QK regime is linearly stable according to the Rayleigh criterion. While magnetic fields can drive turbulence via magnetorotational instabilities~\cite{Balbus03}, it remains unclear whether such instabilities operate in cooler (\ie weakly ionised) discs, where the gas is largely neutral and the ambient magnetism may not play a significant dynamical role. For ordinary (non-conducting) homogeneous fluids, any transition in the QK regime is expected to be subcritical. It is however noted that results from cyclonic subcritical flows, discussed in the previous section, do not offer much guidance for the anti-cyclonic QK regime.} \par
For almost two decades, experiments in the QK regime presented conflicting results. Ji \etal~\cite{Ji2006}, used laser Doppler velocimetry (LDV) to demonstrate that a quiescent flow is maintained even up to $\ReS=\num{2e6}$, which was at the limit of their experimental apparatus. They then surmised by extrapolation that discs were hydrodynamically stable. However, the opposite conclusion was reached by Paoletti \& Lathrop~\cite{Paoletti2011} for slightly lower \ReS, using torque measurements as an integral measure of the angular momentum transport. In both experiments considerable care was taken to minimise end-wall effects. While Ji \etal~\cite{Ji2006} utilised split, independently-rotating end-caps, Paoletti \& Lathrop~\cite{Paoletti2011} measured the torque only on a central portion of the inner cylinder. \par
The influence of end-wall effects has remained an epicentre of discussion and disagreement, but the problem has come into better focus more recently. First, further experiments~\cite{Edlund2014}, where a high-\ReS QK flow was actively perturbed, demonstrated robust stability (\ie laminar flow at high \ReS), bolstering the earlier result~\cite{Ji2006}. Second, direct numerical simulations~\cite{Avila2012, Lopez2017} brought significant clarity to the apparently conflicting experimental results in detailing how boundary effects can be significant even at moderate \ReS. Specifically, if the end-wall conditions are carefully chosen, turbulence becomes confined to thin boundary layers as \ReS increases, while the bulk flow remains laminar~\cite{Lopez2017}. \mod{This result is consistent with axially periodic DNS~\cite{ostilla2014turbulence, shi2017hydrodynamic} and with radially-resolved observations in cyclonic experiments~\cite{Burin2012} and thus points to a general care needed when interpreting results. For additional elaboration on the QK regime and related topics we refer the reader to the papers by Ji \& Goodman~\cite{Ji2023} and by Guseva \& Tobias \cite{Guseva2023}, both in this centennial issue.}

\section{Conclusion}
\label{sec:conc}

Starting with Taylor's groundbreaking work, \TCF has become the archetype to study hydrodynamic stability, bifurcations, pattern formation, transition to turbulence and highly turbulent wall-bounded flows. More generally, the development of the dynamical systems framework to study dissipative systems has been greatly inspired and influenced by experimental observations of this very system. The detailed bifurcation mechanisms and features of the flow patterns depend sensitively on the four dimensionless control parameters of the problem (\Rei, \Reo, \rratio, \aspratio) and on the axial boundary conditions. As a consequence, the dynamical richness of TCF has no limits and in this review we have only summarised briefly some of the main traits. \par
Despite qualitative differences in the instabilities and flow patterns preceding turbulence, the transition to turbulence falls into one of two main categories, which we termed supercritical and subcritical. The first one is characterised by a sequence of bifurcations eventually leading to temporally chaotic dynamics. The ensuing flow patterns fill the whole apparatus and sequentially lose spatial symmetry and coherence. These bifurcations were studied experimentally, numerically and theoretically in great detail in the last century. The catastrophic, subcritical route to turbulence was also studied and characterised experimentally back then, but it is only in the last two decades that an understanding of the underlying mechanisms has emerged. This has been possible thanks to statistically resolved spatio-temporal characterisations of very large systems up to extremely long observation times. The two key physical mechanisms underlying the directed-percolation transition to sustained turbulence in linearly stable TCF are turbulence decay and proliferation. Both are stochastic processes. In the counter-rotating regime, things become even more complicated as the linearly unstable layer close to the inner cylinder interacts with the linearly stable (but nonlinearly unstable) one close to the outer cylinder. This results in a mixture of the two transition routes and is a topic that deserves more attention. It is not only relevant to \TCF, but also to, \eg, curved pipes~\cite{canton2020critical}. \par
A third, perhaps even more intriguing case is the absence of turbulence and hence of a route to \mod{it in the quasi-Keplerian regime. Here, }the weight of the experimental evidence suggests that CCF is nonlinearly stable up to at least $\ReS=\orderof{e6}$. Whether transition occurs at even higher Reynolds numbers remains an outstanding question. The exponential stabilisation of CCF with increasing \ROmega in the cyclonic regime revealed in this review cannot be extrapolated to the anti-cyclonic regime, but highlights the efficiency of the Coriolis force in undermining the regeneration mechanisms of wall-bounded turbulent flows.


\enlargethispage{20pt}


\dataccess{The data presented here will be provided in an online supplement (\pangaea) upon approval for publication. Our code is additionally available here \nsc.}

\aucontribute{D.~Feldmann designed/performed the simulations, analysed the data, and created all figures. M.~Avila designed/coordinated the research and analysed the data. All authors contributed to writing the manuscript and have approved it.}

\competing{The authors declare that they have no competing interests.}

\funding{D.~F. and M.~A. gratefully acknowledge financial support from the German Research Foundation (DFG) through the priority programme \href{http://www.tu-ilmenau.de/turbspp/}{Turbulent Superstructures (SPP1881)} and computational resources provided by the \href{https://www.hlrn.de/?lang=en}{North German Supercomputing Alliance (HLRN)} through  project hbi00041. K.~A. was funded by the Central Research Development Fund of the University of Bremen grant number ZF04B/2019/FB04 (``Independent Project for Postdoc'').}

\ack{The authors would like to thank T.~Kreilos for providing the data presented in Fig.~\ref{fig:kreilosReplot}}


\bibliographystyle{rs.bst}
\bibliography{tcf.bib}

\end{document}